\documentclass[superscriptaddress, 
 reprint,
 amsmath,amssymb,
 aps,
floatfix,
]{revtex4-2}

\usepackage{graphicx}
\usepackage{dcolumn}
\usepackage{hyperref}
\usepackage{braket}
\usepackage[utf8]{inputenc}
\usepackage[english]{babel}
\usepackage{amsmath,calc} 
\usepackage{mathtools} 
\usepackage{ulem}
\usepackage{bm}
\usepackage{floatrow}
\usepackage{todonotes}

\newcommand{\rosso}[1]{{\color{black} #1}}

\newfloatcommand{capbtabbox}{table}[][\FBwidth]

\graphicspath{{./Figures/}}
\begin{document}
\title{An effective density matrix approach for intersubband plasmons coupled to a cavity field: electrical extraction/injection of intersubband polaritons}

\author{M. Lagr{\'e}e}\email{mathurin.lagree@polytechnique.edu}
\affiliation{III-V Lab, Campus Polytechnique, 1, Avenue Augustin Fresnel, RD 128, 91767 Palaiseau cedex, France}
\author{M. Jeannin}
\affiliation{Centre de Nanosciences et de Nanotechnologies (C2N),  CNRS UMR 9001, Universit{\'e} Paris-Saclay, 91120 Palaiseau, France}
\author{G. Quinchard}
\affiliation{III-V Lab, Campus Polytechnique, 1, Avenue Augustin Fresnel, RD 128, 91767 Palaiseau cedex, France}
\author{S. Pes}
\affiliation{III-V Lab, Campus Polytechnique, 1, Avenue Augustin Fresnel, RD 128, 91767 Palaiseau cedex, France}
\author{A. Evirgen}
\affiliation{III-V Lab, Campus Polytechnique, 1, Avenue Augustin Fresnel, RD 128, 91767 Palaiseau cedex, France}
\author{A. Delga}
\affiliation{III-V Lab, Campus Polytechnique, 1, Avenue Augustin Fresnel, RD 128, 91767 Palaiseau cedex, France}
\author{V. Trinit{\'e}}\email{virginie.trinite@3-5lab.fr}
\affiliation{III-V Lab, Campus Polytechnique, 1, Avenue Augustin Fresnel, RD 128, 91767 Palaiseau cedex, France}
\author{R. Colombelli}\email{raffaele.colombelli@c2n.upsaclay.fr}
\affiliation{Centre de Nanosciences et de Nanotechnologies (C2N),  CNRS UMR 9001, Universit{\'e} Paris-Saclay, 91120 Palaiseau, France}


\begin{abstract}
The main technological obstacle hampering the dissemination of modern optoelectronic devices operating with large light-matter coupling strength $\Omega$ is an in-depth comprehension of the carrier current extraction and injection from and into strongly coupled light-matter states, the so-called polaritonic states. 
The main challenge lies in modeling the interaction between excitations of different nature, namely bosonic excitations (the plasmonic ISB excitations) with fermionic excitations (the electrons within the extraction or injection subband). 
In this work, we introduce a comprehensive quantum framework that encompasses both the ISB plasmonic mode and the extractor/injector mode, with a specific emphasis on accurately describing the coherent nature of transport. 
This reveals inherent selection rules dictating the interaction between the ISB plasmon and the extraction/injection subband. 
To incorporate the dynamics of the system, this framework is combined to a density matrix model and a quantum master equation which have the key property to distinguish intra and intersubband mechanisms. 
These theoretical developments are confronted to experimental photocurrent measurements from midinfrared quantum cascade detectors ($\lambda = 10$~\textmu m) embedded in metal-semiconductor-metal microcavities, operating at the onset of the strong light-matter coupling regime ($2\Omega=9.3$ meV). 
We are able to reproduce quantitatively the different features of the photocurrent spectra, notably the relative amplitude evolution of the polaritonic peaks with respect to the voltage bias applied to the structure. 
These results on extraction allow us to elucidate the possibility to effectively inject electronic excitations into ISB plasmonic states, and thus polaritonic states.  
\end{abstract}

\maketitle
\section{Introduction}


The use of electromagnetic resonators like antennas or cavities is an established tool to tailor and improve the properties of optoelectronic devices, whether by increasing the sensitivity, reducing the electronic noise, improving the wall-plug efficiency.
In general, the strategy is to engineer, and typically  increase, the interaction strength between light and an electronic transition in matter.
However, the interaction strength in practical devices is always limited to a small fraction of the photon or electronic transition lifetimes, which places the device in the so-called \textit{weak coupling} regime. 
On the contrary, when the light-matter interaction strength overcomes the losses in the system, the latter enters the {\it strong coupling} regime. 
The new constituents of this system are mixed light-matter states called {\it polaritons}, which can be formed by hybridizing any polarization-carrying matter excitation and a photon field. 

Polariton physics thus emerged as a transverse research field studying the fundamental properties of strongly coupled systems. It revealed a plethora of phenomena, the most recognized being the out-of-equilibrium Bose-Einstein condensation of exciton-polaritons \cite{kasprzak_boseeinstein_2006, bajoni_polariton_2008, carusotto_quantum_2013}.
However, most experiments on polaritons are performed by optical means, whereas practical devices require electrical injection or extraction of charge carriers. Recent experiments sparked new interest in electrical transport in systems under strong light-matter coupling conditions, with the report of increased conductivity in organic molecules \cite{orgiu_conductivity_2015}, or the breakdown of topological protection in quantum Hall systems \cite{appugliese_breakdown_2022,ciuti_cavity_2021}. 
Intense research effort is thus currently devoted to provide an accurate description of transport in systems strongly coupled to a cavity field.
\par
In this  context, intersubband (ISB) polaritons, that originate from the coupling between an intersubband transition in doped semiconductor quantum wells (QW) and a cavity mode, are of particular interest. 
They were first reported in 2003 \cite{dini_microcavity_2003} with absorption experiments,  and that same year electronic detection of the signature of strong coupling was also reported \cite{dupont_vacuumfield_2003}. 
However, proposals for electrical injection and electroluminescence of ISB polariton devices \cite{colombelli_quantum_2005,de2008quantum},  
that were quickly followed by experimental work \cite{sapienza_electrically_2008, jouy_intersubband_2010}, faced the problem of inefficient electrical injection in a polaritonic state. That issue proved insurmountable in the following years \cite{jouy_intersubband_2010,delteil_optical_2011,chastanet_surface_2017,geiser_room_2012}.
To circumvent the problem, the study of the "reverse" process (photo-detection) was proposed to elucidate transport mechanisms in polaritonic ISB electronic devices, with experiments on quantum well infrared photodetectors (QWIP) operating in the strong light-matter coupling regime \cite{vigneron_quantum_2019}. 
%

In this context, we have recently presented a semi-empirical model to describe the electronic photoresponse of quantum cascade detectors (QCD) operating 
in the strong light-matter coupling regime \cite{lagree_direct_2021}. 
Based solely on classical oscillators, it allowed us to shine new light on the polariton-to-electron process, and in particular to conjecture that a direct polariton-to-electron tunnel mechanism may play a major role in such devices. 
This result was obtained at the expense of great simplifications. In particular, because the model is based on classical theory, it cannot include any consideration on the {\it coherence} of the involved processes.

Nevertheless, coherence is of paramount importance when dealing with systems operating in the strong-coupling regime, and even more so for ISB polaritons, that originate from the coupling between a cavity mode and a collective excitation. 
ISB transitions, that are more rigorously defined as ISB plasmons\cite{ando_electronic_1982, helm_intersubband_1999,delteil_charge_2012}, are collective matter excitations originating from the electronic plasma inside a semiconductor quantum well, subject to its own Coulomb interaction.
This is in stark contrast to, for instance, exciton-polaritons that result from an ensemble of single-particle transitions. 
The main consequence is the presence of \textit{dark states}, that do not couple to the electromagnetic field, but do participate in electronic transport. This has important consequences on the behavior of ISB polariton systems under electrical injection. 

\begin{center}
\begin{figure}
\includegraphics[width=\linewidth]{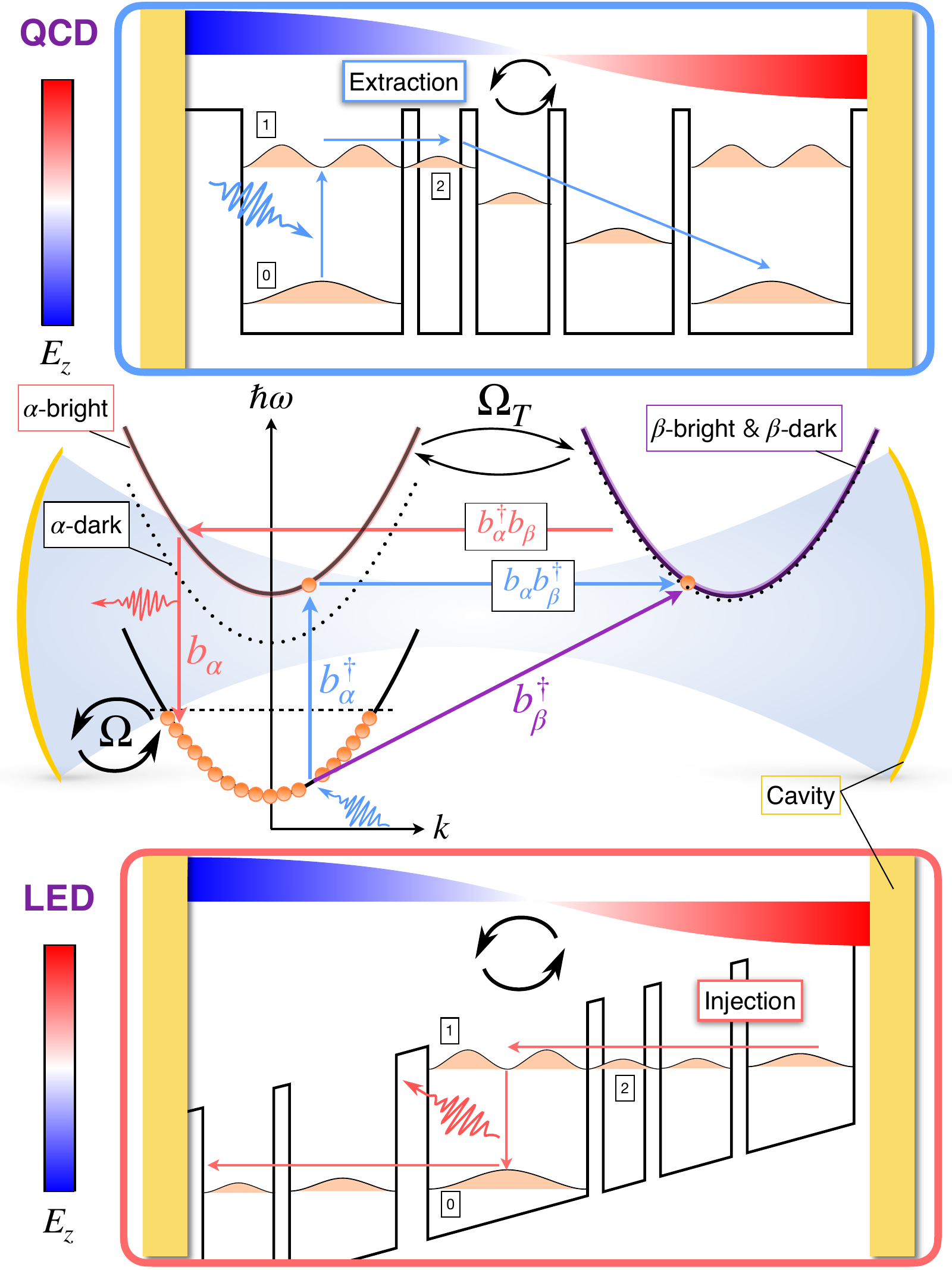}
\caption{(center) Schematic representation of the system in momentum space. Bright and dark states are represented for both $\alpha$ (0$\rightarrow$1) and $\beta$ (0$\rightarrow$2) transitions. Note that the $\beta$-bright state is degenerated with the $\beta$-dark states. The different important operators and their effect on the transport of the excitations are represented. The blue path represents a detection process, whereas the red path represents an injection process. (Top) Typical bandstructure for a quantum cascade \textit{detector}. The main \textit{extraction} pathway is represented in blue. (Bottom) Typical bandstructure for a quantum cascade \textit{emitter}. The main \textit{injection} pathway is represente in red. The cavity electric field $E_z$ is also schematically superimposed on the figure. }
\label{fig:system}
\end{figure}
\end{center}

In this paper, we propose a quantum description of QCDs based on a density matrix formalism, that we compare to a complete set of experimental data. 
Crucially, this approach allows us to describe (de)coherence and dissipation in the system. Our goal is to develop a theoretical description that permits to explain the \textit{electronic extraction} process (photo-detection), and that - at the same time - provides a more suitable vantage point to elucidate the more complex \textit{electronic injection} process leading to light emission. 
\rosso{We note that a very recent work reports experimental results and proposes an alternative transport model for similar QCD structures operating in the strong coupling regime~\cite{pisani_electronic_2023}. It works explicitly with the Fermionic approach, without performing the bozonisation steps. 
While similar conclusions are drawn in the photo-detection case, the work we present raises fundamental open questions and presents ways forward to the case of electrically pumped polaritonic light emitters.}

In the first part, we develop the model and derive the main observable quantities, notably the photocurrent generated  by an exciting external photon field.\\ 
In the second part, we validate the theoretical results by studying the photoresponse of quantum cascade detectors operating 
in the strong coupling regime as a function of the applied bias. 
We compare the values obtained in our model 
with a in-house code based on Ref. \cite{trinite2011modelling} that models the electronic transport in a more rigorous way, but does not 
incorporate the cavity effects \cite{koeniguer2006electronic,buffaz2010role}.\\
In the last part, we discuss the implications of the main assumption at the basis of our new model, and extend them to the case of electrical injection.

The system under study is sketched in the central part of Fig.~\ref{fig:system}. 
It consists of two electronic subbands confined inside a QW, here represented in momentum space. 
The second subband is tunnel-coupled to the fundamental state of an adjacent QW, and the whole system is embedded inside a cavity. 
The system can operate as a detector, acting as a QCD (top sketch), when it is excited by a photon that generates a photocurrent. This path is represented by blue arrows. 
It is also possible to inject electrons in the system (red arrows and bottom sketch), when an electric bias is applied, that can eventually lead to photon emission. In this case the device behaves as a polaritonic LED.

%
%
\section{An effective density matrix approach for electronic transport in cavity-coupled QCDs \label{section:theory}}
\subsection{Bosonization of the active optical transition}
%
We start by defining the annihilation and creation operators $c_{\lambda \mathbf{k}}$ and $c_{\lambda \mathbf{k}}^\dagger$, the fermionic operators related to the creation and annihilation of electrons in subbands $\lambda = \{0,1,2\}$ (see Fig. \ref{fig:system}). We impose $T=0$K and we assume that all $N$ electrons are contained inside the $0$-subband without external excitation. The one-particle quantum state $|1,\mathbf{k}\rangle$ of electronic wave vector $\mathbf{k}$, representing a state where one electron is in subband $\lambda=1$ is:
\begin{equation}
    \ket{1,\bf{k}}= c_{1\bf{k}}^\dagger c_{0\bf{k}}|F\rangle \label{Eq:oneParticle}
\end{equation}
where $|F\rangle$ denotes the fundamental Fermi state (equilibrium state, where all the electrons are contained in subband $\lambda=0$).
For now, we restrain the problem to the $\lambda=0,1$ subbands, that form the intersubband optical transition. This transition will be denoted as $\alpha$. Following the developments of Ref. \cite{todorov2012intersubband}, to describe the photo-excitation of an electron in the $\alpha$-transition, it is relevant to switch from the fermionic basis formed by the $\ket{1,\bf{k}}$ states to a new basis of states $\{|B_i^\alpha \rangle \}_{i=[1:N]}$.

We have:
\begin{equation}
\ket{B_i^\alpha} = \sum_{ |\mathbf{k}|< \mathbf{k}_F} w_{i \mathbf{k}}^{\alpha} \ket{1,\mathbf{k} }
\end{equation}
Since the system is considered at $T=0$ K, only $|\mathbf{k}|<\mathbf{k}_F$ states are occupied, $\mathbf{k}_F$ the module of the $k$-wavevector corresponding to the Fermi level of 0-subband. The $\{|B_i^\alpha \rangle \}_{i=[1:N]}$ basis only covers the single-excitation subspace (only one photo-excited electron per-subband), which is sufficient in the case of a weak excitation regime. The coefficients $w_{i \mathbf{k}}^{\alpha}$ are defined as:
\begin{eqnarray}
w_{1\mathbf{k}}^{\alpha} & =&  \frac{1}{\sqrt{N}} ~~~ \forall \mathbf{k} \label{Eq:bright_coeff_alpha}\\
\sum_\mathbf{k} w_{i \mathbf{k}}^{\alpha} & =& 0  ~~~ \forall i\neq 1  \label{Eq:dark_coeff_alpha}
\end{eqnarray}
The $|B_1^\alpha\rangle$ state, of eigenenergy equal to the ISB transition energy $\omega_\alpha = \omega_{1}-\omega_0$ (assuming parabolic dispersion), has the remarkable property of holding \textit{the entire oscillator strength} of the $\alpha$ transition:
\begin{equation}
\langle F | \hat{d} | B_1^\alpha \rangle = z_\alpha \sqrt{N}
\end{equation}
where  $\hat{d}$ denotes the dipole operator and $z_\alpha$ the dipole strength of one electronic transition. The $|B_1^\alpha\rangle$ state is called the \textit{bright} state: it is formed by the \textit{coherent} superposition of the one-particle fermionic states $|1,\mathbf{k}\rangle$ of the $\alpha$-transition and it holds the entire capacity of light-matter interaction. 
The $\{|B_i^\alpha \rangle \}_{i=[2:N]}$ are called the \textit{dark} states since they can not interact with the light:
\begin{equation}
\langle F | \hat{d} | B_i^\alpha \rangle = 0 ~~~ i\neq 1
\end{equation}
From these developments, one can define bright state destruction and creation operators $b_\alpha$ and $b_\alpha^\dagger$ which describe the collective excitation of the $\alpha$-transition:
\begin{equation}
b_\alpha^{\dagger}= \frac{1}{\sqrt{N}} \sum_\mathbf{k} c_{1 \mathbf{k} }^{\dagger} c_{0 \mathbf{k} }
\end{equation}
In a weak excitation regime and for a large number of electrons $N$, $b_\alpha$ can be approximated as a bosonic operator. $b_\alpha$ and $b_\alpha^\dagger$ respectively demote and promote excitations inside the bright state $|B_1^\alpha\rangle$. 

The final step in this development is to include the plasmonic effect $\omega_P$ of the electronic polarizations. 
The diagonalization of the plasmonic Hamiltonian leads to the emergence of new operators of eigen-energy $\tilde{\omega}_\alpha= \sqrt{\omega_\alpha^2 + \omega_P^2 }$ and a plasmonic bright state that is still orthogonal to the dark states \cite{todorov2012intersubband}. 
Mathematically, this new state is essentially the same as the previous bright state, except that it is no longer degenerated with the dark states: for simplicity, we will keep the notation $|B_1^\alpha\rangle$ and $b_\alpha$ for respectively the bright state and the corresponding creation operator. 
Note: at this stage we did not introduce strong light-matter coupling yet. This derivation is therefore valid in any coupling regime. 

\subsection{Bosonization of the extractor: the tunnel-coupling Hamiltonian}
\label{sec:extractor}

%
We now turn to the insertion of the extraction subband in the formalism. As outlined in Refs. \cite{de2008quantum,de2009quantum}, the mixing of bosonic (the plasmonic ISB excitations) and fermionic (the electrons in the extraction subband) degrees of freedom is necessary to correctly model the transport mechanisms that take place in an optically excited ISB system.
The focus of our paper is on ISB systems strongly coupled to a photonic mode, but we stress that the above consideration is valid also in the weak-coupling regime.
When a photon is absorbed by an ISB transition, it generates a bosonic excitation: an ISB plasmon. But the measured current, in a detector, is of course of fermionic nature.

In the case where the extraction subband is explicitly included in the system dynamics (and not only in the form of an external bath), it becomes an extremely tedious task to keep track of all these degrees of freedom. Effectively, one correct way to describe the interaction between these excitations of different nature is to use a full fermionic Hamiltonian of extremely large dimension. It is a significant mathematical challenge that demands considerable effort, and the nature of transport cannot be straightforwardly interpreted due to this complexity. 

%
%
In this work, we overcome this strong limitation with a key modification: we propose to depict the subband $\lambda=2$ with a bosonic operator in the context of an extraction process.
This approach has several advantages, and - as we will discuss later on - it might also permit to address the scenario involving an injection process. 
To explicitly incorporate subband $\lambda=2$ into our formalism, we  introduce the one-particle fermionic states $|2,\mathbf{k}\rangle$ of the $\beta$-transition:
\begin{eqnarray}
    |2,\mathbf{k}\rangle = c_{2\mathbf{k}}^\dagger c_{0\mathbf{k}}|F\rangle
\end{eqnarray}
Analogous to the $\alpha$-transition, we will not use this fermionic state basis and instead employ a new ortho-normal basis $\{|B_i^\beta \rangle \}_{i=[1:N]}$ defined as:
\begin{equation}
\ket{B_i^\beta} = \sum_{ |\mathbf{k}|< \mathbf{k}_f} w_{i \mathbf{k}}^{\beta} \ket{2,\mathbf{k} } \label{Eq:basis_beta}
\end{equation}
where the coefficients $w_{i \mathbf{k}}^{\beta}$ are chosen such that: 
\begin{eqnarray}
w_{1\mathbf{k}}^{\beta} & =&  \frac{1}{\sqrt{N}} ~~~ \forall k \label{Eq:bright_coeff_beta}\\
\sum_k w_{i  \mathbf{k}}^{\beta} & =& 0  ~~~ \forall i \neq 1  \label{Eq:dark_coeff_beta}
\end{eqnarray}
The construction of this basis follows a similar approach as that of the $\{|B_i^\alpha \rangle \}_{i=[1:N]}$ basis. Specifically, the first state $|B_1^\beta\rangle$ is the bright state of the $\beta$-transition, while the remaining states $\{|B_i^\beta \rangle \}_{i=[2:N]}$ are the dark states of this same transition. 
However, this time, the oscillator strength of a diagonal transition being very small, we have $z_\beta\ll z_\alpha$ and thus the bright and dark states of the extractor are degenerated. 
Note that the one excitation subspace describing subband $1$ and $2$, of dimension $2N$, is spanned by the concatenation of the $\{|B_i^\alpha \rangle \}_{i=[1:N]}$ and $\{|B_i^\beta \rangle \}_{i=[1:N]}$ basis.

The introduction of this new basis is valuable to evaluate the tunnel coupling between subbands 1 and 2 within the regime of strong light-matter coupling. The tunnel coupling operator $\hat{T}$ can be defined as:
\begin{eqnarray}
    \hat{T} = \Omega_T \sum_\mathbf{k} (c_{2\mathbf{k}} c_{1\mathbf{k}}^\dagger + c_{2\mathbf{k}}^\dagger c_{1\mathbf{k}} )
\end{eqnarray}
where $\Omega_T$ is the tunnel coupling strength. Using equations \eqref{Eq:bright_coeff_alpha}, \eqref{Eq:dark_coeff_alpha}, \eqref{Eq:bright_coeff_beta} and \eqref{Eq:dark_coeff_beta}, we compute the tunnel interaction between subbands $1$ and $2$:
\begin{align}
\braket{B_1^\alpha | \hat{T} | B_1^\beta} = & \Omega_T &\label{Eq:tunnel_brightbright}  \\
\braket{B_1^\alpha | \hat{T} | B_j^\beta} = & 0  &  j \neq 1  \label{Eq:tunnel_brightdark}\\
\braket{B_i^\alpha | \hat{T} | B_1^\beta} = & 0  & i \neq 1 \label{Eq:tunnel_darkbright} \\
\braket{B_i^\alpha | \hat{T} | B_j^\beta} = & \Omega_T \sum_\mathbf{k} w_{i\mathbf{k}}^{\alpha \ast} w_{j\mathbf{k}}^\beta 
                                      & i \neq 1, j \neq 1 \label{Eq:tunnel_darkdark}
\end{align}
The above relations, that are \textit{de facto} selection rules, are one of the key results of this work:
through tunnel interaction, it is not possible to transition from a dark state to a bright state (Eq. \eqref{Eq:tunnel_brightdark}) or vice versa (Eq. \eqref{Eq:tunnel_darkbright}). 
Obviously, dark states can interact with each other through tunnel coupling (Eq. \eqref{Eq:tunnel_darkdark}), and the same applies to bright states as well (Eq. \eqref{Eq:tunnel_brightbright}). 

These results have crucial implications on the nature of electronic transport in a QCD. 
For a detection process, where light promotes excitations into the $|B_1^\alpha\rangle$ bright state, the previous results suggest that an optical excitation can generate an electronic current in only two ways:
\begin{enumerate}
    \item  Direct tunnelling into the extractor bright state $|B_1^\beta\rangle$, preserving the coherent nature of the excitation, and subsequent decay - with loss of coherence - into an extractor dark state $|B_{i\neq1}^\beta\rangle$

or
    
 \item First decay - with loss of coherence - into an ISB dark state $|B_{i\neq1}^\alpha\rangle$ in the active region, and subsequent tunneling into an extractor dark state $|B_{i\neq1}^\beta\rangle$ 
\end{enumerate}

Other channels involving bright-to-dark tunneling should not be considered, as they are prohibited by selection rules \eqref{Eq:tunnel_brightdark}\eqref{Eq:tunnel_darkbright}.
Once in the extractor dark states, the electronic excitation will simply decay in the remaining cascade, generating photocurrent.  
We stress that the construction of the new $\beta$ basis merely extended the procedure applied to the $\alpha$ transition (detailed in reference \cite{todorov2012intersubband}) to the $\beta$ transition, without additional hypothesis. By implementing this basis transformation, the comprehension of the transport process is streamlined, leading to the natural emergence of the selection rules presented in Equation \eqref{Eq:tunnel_brightbright} to \eqref{Eq:tunnel_darkdark}. 
In the following section, we will assess the need to actually incorporate the dark states from both the $\alpha$ and $\beta$-transitions to replicate the experimental photocurrent measurements from a QCD operating in the strong light-matter coupling regime.
The implications of this section for an electronic injection process into polaritonic states will be discussed in section \ref{section:injection}.
%

%

\subsection{Introducing dissipation and decoherence in the model \label{section:bosonization}}
In the following, we develop an effective density matrix model of the photocurrent extraction. 
We apply a drastic choice in the description of the system: we limit the extraction model to the transport induced by the bright states $ |B_1^\alpha \rangle$ and $| B_1^\beta \rangle$. 
The dark states from both the $\alpha$ and $\beta$-transitions are omitted. Both subbands $1$ and $2$ will thus be described only using bosonic operators. 
This is equivalent to choose scenario (1) among the two described at the end of the previous section: 
direct tunnelling into the extractor bright state $|B_1^\beta\rangle$ (preserving the coherent nature of the excitation), and subsequent decay - with loss of coherence - into an extractor dark state $|B_{i\neq1}^\eta\rangle$.
This choice was already implicit in the approach that we have employed in our previous work based on a {\it classical} description of the electronic transport, using coupled mode theory \cite{lagree_direct_2021}. 

We now go beyond this classical model using a quantum master equation. The key addition is the introduction of \textit{decoherence} in the system, that is distinct from dissipation. 
%

%
In terms of spectral effects, decoherence impacts the broadening of the photocurrent peaks, while dissipation primarily affects their amplitude. 
In the experimental study we will report in Sec.~\ref{sec:experiment}, bias will be varied, and - as a result - the amplitude of the peaks will be affected more than their broadening. It will be essential to differentiate between the effects of decoherence and dissipation, a distinction that was previously impossible to achieve with the classical model. 

We define the operator $b_\beta$ using our new basis from equations \eqref{Eq:basis_beta} and \eqref{Eq:bright_coeff_beta}:
\begin{eqnarray}
    b^{\dagger}_{\beta} &=& \frac{1}{\sqrt{N}} \sum_\mathbf{k} c_{2 \mathbf{k}}^{\dagger} c_{0\mathbf{k}}\\
    b^\dagger_\beta |F\rangle &=& |B_1^\beta\rangle  
\end{eqnarray}
Using the fermionic commutation rules and a weak excitation regime, we have:
\begin{eqnarray}
    [b_\beta,b_\beta^\dagger] = \frac{\hat{N}_0 - \hat{N}_2}{N} \approx \hat{\mathcal{I}_d}
\end{eqnarray}
where $\hat{N}_i$ is the population operator of subband $i$ and $\hat{\mathcal{I}_d}$ the identity operator.  
$b_\beta$ can thus be approximated as a bosonic operator: $b_\beta$ and $b_\beta^\dagger$ describe the destruction and creation of electronic excitations inside the extraction mode, of eigen-frequency $\omega_\beta = \omega_{2} - \omega_{0}$. The related Hamiltonian is:
\begin{equation}
    \hat{\mathcal{H}}_\beta = \omega_\beta b_\beta^\dagger b_\beta
\end{equation}
We restrict the tunnel interaction to the interaction between the plasmonic bright mode and this new extraction mode. This drastically simplifies the tunnel interaction Hamiltonian described in Eq. \eqref{Eq:tunnel_brightbright}. 
The restricted Hamiltonian $\hat{T}_\text{bright}$ is:
\begin{equation}
    \hat{T}_\text{bright} = \Omega_T (b_\alpha^\dagger b_\beta +  b_\alpha b_\beta^\dagger)
\end{equation}
The TM$_{01}$ electromagnetic mode confined in the patch antennas will be modeled as a standard optical resonator of frequency $\omega_c$, using $a_c$ and $a_c^\dagger$ bosonic destruction and creation operators. Using the rotating wave approximation to describe the light-matter interaction, the time dependent Hamiltonian $\mathcal{H}(t)$ of the whole system reads:
\begin{eqnarray}
    \hat{\mathcal{H}}(t) & = &  \omega_c a_c^{\dagger} a_c  + \tilde{\omega}_\alpha b_\alpha^{\dagger} b_\alpha + \omega_\beta b_\beta^{\dagger} b_\beta \nonumber \\
                   & & + \Omega \left( a_c^{\dagger} b_\alpha + a_c b_\alpha^{\dagger} \right) 
                    + \Omega_T \left( b_\alpha^\dagger b_\beta +  b_\alpha b^\dagger_\beta \right) \\
                   & & + \kappa_c s_+ \left( a_c e^{i\omega t} + a_c e^{-i \omega t} \right) \nonumber
\end{eqnarray}
where $s_+$ is the amplitude of the incoming light excitation, $\omega$ its frequency, and $\kappa_c$ is the coupling constant between this external field and the confined optical mode inside the cavity. 

We map this system on an equivalent open quantum system described by the reduced density matrix $\rho$. Under standard Born-Markov approximations, the time evolution of the density matrix $\rho$ obey the following quantum master equation \cite{breuer2002theory} ($\hbar=1$ for clarity):
    \begin{eqnarray}
\frac{\text{d}\rho(t)}{\text{d}t} & =& - i \big{[} \mathcal{H}(t),\rho \big{]} \label{Eq:QMMEeq} \\
     && + \gamma_\alpha \mathcal{L}\left[b_\alpha, \rho\right] 
        + \gamma_\beta \mathcal{L}\left[b_\beta, \rho \right] 
        + (\gamma_c+\Gamma_c) \mathcal{L}\left[a_c, \rho \right]
          \nonumber \\
     && + \gamma_\alpha^\text{intra} \mathcal{L}\left[b_\alpha^{\dagger}b_\alpha, \rho \right]  
        + \gamma_\beta^\text{intra} \mathcal{L}\left[b_\beta^{\dagger}b_\beta, \rho \right]  \nonumber 
\end{eqnarray}
where the $\mathcal{L}$ are Lindblad super-operators modeling the dissipative and decoherent interactions of the environment with the system. For any operator $\hat{A}$, a super-operator $\mathcal{L}$ reads:
\begin{eqnarray}
    \mathcal{L}[\hat{A},\rho] = 2 \hat{A} \rho \hat{A}^\dagger - (\hat{A}^\dagger \hat{A} \rho + \rho \hat{A}^\dagger \hat{A})
\end{eqnarray}
The plasmonic ISB excitations are mainly dissipated through their interaction with interface roughness, at a non-radiative rate $\gamma_\alpha$. Similarly, the extractor dissipates electrons into the next period at a non-radiative rate $\gamma_\beta$, and is responsible for the generation of electrical current inside the structure. $\gamma_\beta$ represents an effective dissipation rate that takes into consideration the remaining electronic cascade.  The cavity also dissipates photons (mainly through undesired free-carriers absorption) at a rate $\gamma_c$, but also through a spontaneous emission channel, at a radiative rate $\Gamma_c$. Note that the radiative coupling $\kappa_c$ is related to the radiative damping through $\kappa_c = \sqrt{2 \Gamma_c}$ \cite{suh2004temporal}. 

The main difference with our previous work \cite{lagree_direct_2021} lies in the ability to explicitly introduce the \textit{intra}-subband scattering through the pure decoherence terms $\gamma_\alpha^\text{intra} \mathcal{L}\left[b_\alpha^{\dagger}b_\alpha, \rho \right]$ (resp. $\gamma_\beta^\text{intra} \mathcal{L}[b_\beta^{\dagger}b_\beta, \rho ]$) \cite{schlosshauer2007decoherence}. These terms model pure decoherence damping without excitation dissipation (the intra-subband scattering thermalize excitations inside a subband without dissipating them into an other subband). By using the density matrix formalism, it thus becomes possible to differentiate between the effects of inter-subband (dissipation) and intra-subband (pure decoherence) processes on the evolution of the system (and ultimately on the shape of the calculated photoresponse spectra). More details on the necessity to distinguish intra and intersubband scatterings can be found in Appendix \ref{section:transferFunction}.

\subsection{Deriving observable quantities for comparison with experiments}
Equation \ref{Eq:QMMEeq} can be solved numerically in steady state. The solution is a stationnary reduced density matrix $\rho_S$, and any observable $\hat{O}$ can then be computed using:
\begin{eqnarray}
    \langle \hat{O} \rangle = \text{Tr}(\hat{O} \rho_s)
\end{eqnarray}
where $\text{Tr}$ represents the trace function. We can then compute the different interesting quantities of the system. The system total absorption is the sum of the power dissipated into the different decay channels, normalized by the incoming power $|s_+|^2$:
\begin{eqnarray}  
\mathcal{A}_\text{tot} &=&  \mathcal{A}_c + \mathcal{A}_\alpha  + \mathcal{A}_\beta  \\
                       &=& 2 \gamma_c \frac{\braket{a_c^{\dagger} a_c}}{\left| s^+ \right|^2} +  2 \gamma_{\alpha} \frac{\braket{b_\alpha^{\dagger} b_\alpha}}{\left| s^+ \right|^2} + 2 \gamma_{\beta} \frac{\braket{b_\beta^{\dagger} b_\beta}}{\left| s^+ \right|^2} 
\end{eqnarray}
where $\mathcal{A}_c$, $\mathcal{A}_\alpha$ and $\mathcal{A}_\beta$ represent respectively the cavity, ISB and extraction absorptions. 

The net photocurrent $\mathcal{J}_\beta$ is defined as the current under illumination. $\mathcal{J}_\beta$ is proportional to the power dissipated from a period to the next adjacent period. This is exactly the power dissipated by the extraction mode $\beta$:
\begin{eqnarray}
\mathcal{J}_\beta = 2 \gamma_{\beta} \braket{b_\beta^{\dagger} b_\beta} \label{Eq:J_beta}
\end{eqnarray}

Note: this is a phenomenological interpretation of the photocurrent. It is in fact expected that an excitation inside the bright extractor state $|B_1^\beta\rangle$ should first decay in the dark states $|B_{i\neq 1}^\beta\rangle$ before being extracted in the electronic cascade and contribute to the photocurrent. 
We choose to neglect these dark extractor states such that the power is directly dissipated from the bright extractor state. This also applies on the ISB dissipation, where the $|B_{i\neq 1}^\alpha\rangle$ dark states are neglected when considering the non-radiative dissipation $\gamma_\alpha$. 

\section{Experimental validation in photo-detection: the polariton-to-current process}
\label{sec:experiment}

\subsection{Experimental details}

The samples investigated in this study are the same as those already studied in Ref. \cite{lagree_direct_2021}.  
They are processed into 8 $\times$ 8 (approximately equal to 50 $\times$ 50 \textmu m$^2$) patch antenna arrays, with the patches connected through 250-nm thin metallic wires (see Fig. \ref{fig:patchs} in Appendix \ref{section:expsystem}). Details of the processing can be found in \cite{quinchard2022high}.
The samples are cooled down to T = 78K in a cryostat, and they are illuminated by light from a globar source at normal incidence.
The photocurrent spectra are acquired in rapid scan mode, after amplification using a low-noise transimpedance amplifier.

We extend the data presented in \cite{lagree_direct_2021}, and now present measurements with voltage bias applied to the samples.  
The applied electric field ranges from $F=-25~$kV.cm$^{-1}$ to $F=8~$kV.cm$^{-1}$. 
We have fabricated several array designs ($p$, $s$), with $p$ the inter-patch period of the array, and $s$ the lateral dimensions of the patches.
However, to allow for a quantitative comparison, we present measurements under an applied electric field for two samples only, with same $p=$7 \textmu m, and $s$= 1.5 \textmu m and $s$=1.55 \textmu m, respectively, as reported in Fig.~\ref{fig:Exp_Photocurrent} (continuous lines).
Additional measurements can be found in Appendix \ref{section:photocurrent_additional}. While the relative amplitude of the spectra when varying the bias contains meaningful information of the electronic transport, one should exercise caution when comparing the amplitudes of different pairs ($p$, $s$) as the experimental protocol does not ensure a consistent illumination between each measurement of the device. 

\begin{figure*}[!ht]
    \centering
    \includegraphics[width=14cm]{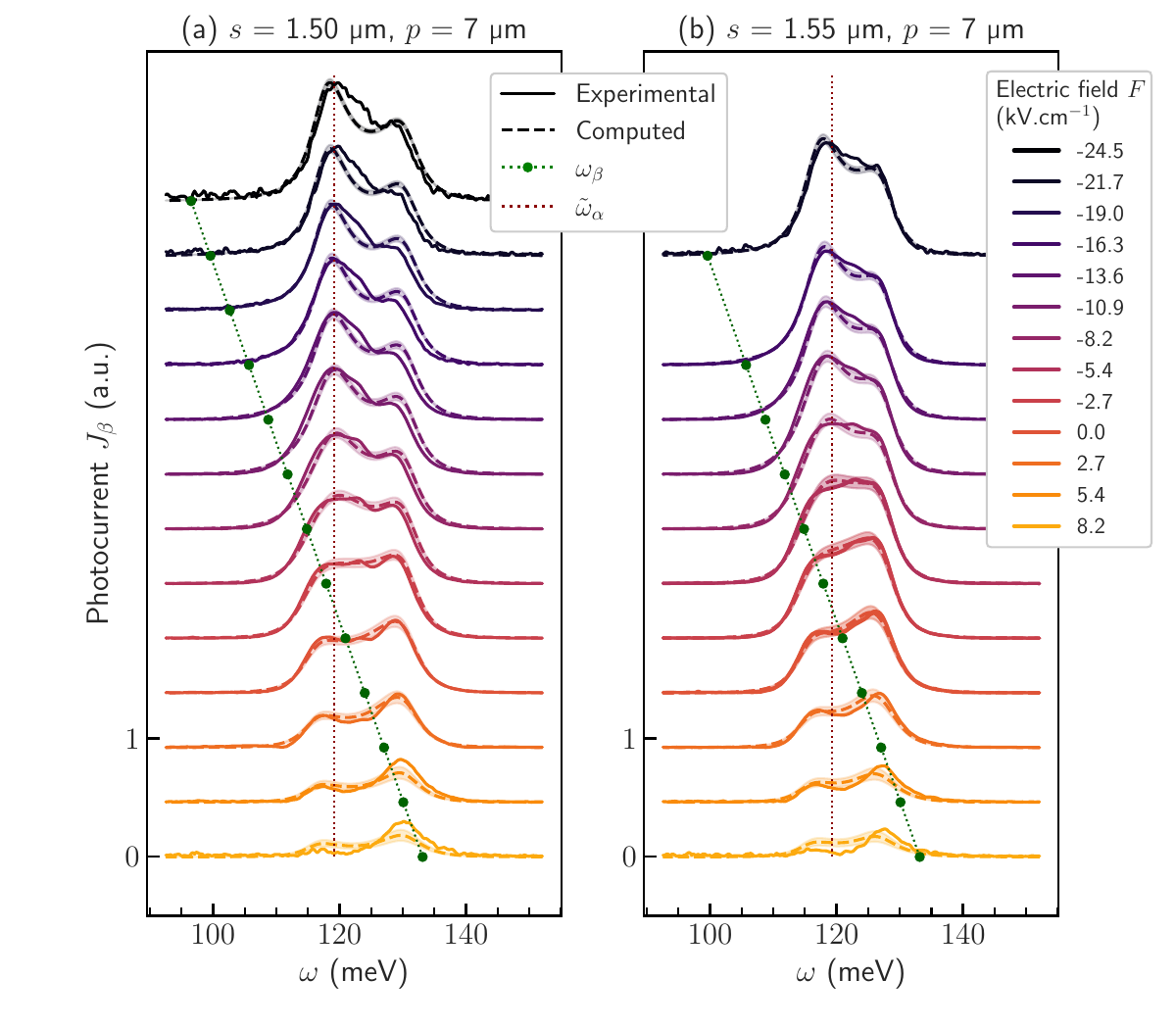}
    \caption{Normalized photocurrent measurements (continuous lines) and quantum master equation global fit (dashed lines), for two cavity geometries [a] $s=1.50$ \textmu m, $p=7$ \textmu m and [b] $s=1.55$ \textmu m, $p=7$ \textmu m. Offsets are added for clarity. Filled areas represent the errors of the fit parameters propagated onto the spectra. The extractor frequency $\omega_\beta(F)$, dependant of the electric field $F$, and the plasma-shifted ISB transition $\tilde{\omega}_\alpha$ are both superimposed on the spectra. Additional results can be found in Appendix \ref{section:photocurrent_additional}. }
    \label{fig:Exp_Photocurrent}
\end{figure*}
Two photocurrent peaks are clearly visible in Fig.~\ref{fig:Exp_Photocurrent}, signature of the strong light-matter coupling regime. Note: the peaks under consideration cannot be confused with the two peaks arising from coupled subbands (tunnel coupling), since the peak positions would change with the applied bias in the latter case. Here, the energy splitting  (for a given pair $p$, $s$) is constant regardless of the applied field. 
For all  ($p$, $s$) couples studied, the global amplitude of the photocurrent spectra evolves with the applied electric field $F$. 
A maximum amplitude is observed around $F=-10$ kV.cm$^{-1}$. The noise level increases strongly when the absolute amplitude of the field $|F|$ increases. The noise level is the direct consequence of the increase of the parasitic dark current with the electric field and - as is well known~\cite{delga2012master,hakl_ultrafast_2021} - it affects the range of exploitable field $F$ for device applications. 

The relative amplitude of these peaks inverts with respect to the applied field $F$, with the equal amplitude condition of the two polaritonic photo-detection peak found for a negative field $F \approx -5$ kV.cm$^{-1}$. Below this threshold, the low energy peak dominates. Inversely, for $F > -5$kV.cm$^{-1}$, it is the high energy peak that dominates. This phenomenon can be attributed to the realignment of the subbands under the influence of the applied bias. When a highly negative voltage is applied, the subbands follow a clear staircase structure
(see Fig. \ref{fig:BandStructure} in Appendix \ref{section:expsystem} for the QCD bandstructure), which facilitates the extraction process. Conversely, at positive voltages, the subband cascade becomes less organized, hindering the extraction process.

%

%
\subsection{System parameters and constraints}
Before applying the theoretical developments of section \ref{section:theory} to the experimental data, let us detail the system parameters and the constraints applied to them.

The photonic degrees of freedom are the cavity parameters $\omega_c$, $\gamma_c$ and $\Gamma_c$, that are independent of the applied electric field $F$. 
They only depend on the geometrical parameters ($p$, $s$) of the cavities~\cite{todorov2010optical,balanis2016antenna,palaferri2018antenna}:
\begin{eqnarray}
    \omega_c(s) &=& \frac{\pi c_0}{n_\text{eff} s} \\
    \Gamma_c(p) &=& \frac{\alpha_c}{p^2} 
\end{eqnarray}
where $c_0$ is the light velocity, $n_\text{eff}$ is the effective index of the cavity, that represents the effective medium composed of the semiconductor contacts and of the undoped periodic structure embedded between the gold layers forming that cavity, and $\alpha_c$ is the cavity dispersion loss factor. 
We choose to constrain $n_\text{eff}$, $\alpha_c$ and $\gamma_c$ to the values obtained from our prior investigation of the same samples~\cite{lagree_direct_2021}, where the photocurrent of several samples with different ($s$,$p$) couples have been studied for $F=0$ kV.cm$^{-1}$:
\begin{eqnarray}
    n_\text{eff} &=& 3.22  \\
    \alpha_c &=& 29.1 ~~\text{meV.\textmu m$^2$} \\
    \gamma_c &=& 3.4 ~~\text{meV}
\end{eqnarray}
The cavity parameters are thus excluded from the fitting process.

Several electronic degree of freedom can also be fixed or constrained independently of our density matrix model. 
%
The parameters of the ISB transition in the active QW ($\alpha$) are assumed independent of the applied electric field $F$: the transition is vertical in a single quantum well and therefore is very marginally affected by the applied bias. The ISB frequency $\omega_\alpha$ and the plasma frequency $\omega_P$ could be computed from our sequential transport software~\cite{trinite2011modelling}. However, it is common to observe disparities between expected and measured doping levels (up to 15\%). Experimental discrepancies also affect the ISB frequency (up to 5\%), usually caused by the quality of the quantum wells interfaces during the epitaxial process. To account for these disparities, and since both $\omega_\alpha$ and $\omega_P$ are crucial parameters to reproduce the strong coupling measurements, we chose to let these parameters free during the fitting process:
\begin{eqnarray}
    \tilde{\omega}_\alpha &=& \sqrt{\omega_\alpha^2 + \omega_P^2}
\end{eqnarray}
Note: the light-matter coupling constant $\Omega$ is parametrized using $\omega_P$:
\begin{eqnarray}
    \Omega= \frac{\omega_P}{2}\sqrt{f_w}
\end{eqnarray}
with $f_w$ ($\approx$ 0.17), the computed overlap factor between the cavity field and the doped active quantum wells.
\\
Two additional $\alpha$ parameters can be computed using our sequential transport software: the non-radiative dissipation rate $\gamma_\alpha$ of the $\alpha$ plasmon from the excited subband to the fundamental subband, and the tunnel coupling $\Omega_T$. 
We compute $\gamma_\alpha = 0.66$~meV and $\Omega_T = 4.2$~meV, respectively. The new parameter of our transport model in the strong coupling regime, the intra-subband rate $\gamma_\alpha^\text{intra}$, will instead be fitted.

The parameters related to the extractor $\beta$ are instead dependent on the electric field $F$: the extractor energy shifts with respect to the upper excited state of the ISB transition when a bias is applied to the structure. 
The misalignment is approximated as linear:
\begin{eqnarray}
    \omega_\beta (F) = \alpha_F F + \omega_\beta^0
\end{eqnarray}
where $\alpha_F$ is the linear coefficient and $\omega_\beta^0$ is the extractor energy for $F=0$. This dispersion can be computed using our sequential transport software and is injected into the model:
\begin{eqnarray}
    \alpha_F &=& 1.12 ~~\text{meV/(kV.cm$^{-1}$)}\\
    \omega_\beta^0 &=& 124 ~~ \text{meV}
\end{eqnarray}
Similarly to $\gamma_\alpha^\text{intra}$, $\gamma_\beta^\text{intra}$ will be a fitting parameter common to the whole data set.
\\
Finally, we expect the misalignment of the cascade with the electric field to modify the value of the effective extraction rate $\gamma_\beta(F)$. 
$\gamma_\beta$ is one of the most important parameters of the fitting process, as it controls the relative amplitude of the spectra. Although we suspect that it might closely match with the actual extraction rate calculated from our sequential transport model, we decided to keep it as a free parameter: for each measured electric field value $F_i$, we fit one extraction rate $\gamma_\beta(F_i)$. Note: $\gamma_\beta(F_i)$ is independent of the geometrical parameters $p$ and $s$. In summary, $\omega_\alpha$, $\omega_P$ and $\gamma_\alpha^\text{intra}$ and $\gamma_\beta^\text{intra}$ are fitting parameters common to the whole data set, and their initial values for the fit will be based on the ones derived by our software. 

\subsection{Discussion on the validity of the fit}
In this section, we perform a global fit on the whole experimental photocurrent dataset (Fig.~\ref{fig:Exp_Photocurrent}), using the parameters constraints exposed in the previous section. We solve  Eq.\eqref{Eq:QMMEeq} in the stationary regime (using the QuTiP python library \cite{johansson2012qutip}) to evaluate the theoretical photocurrent $J_\beta$, as per Eq.\eqref{Eq:J_beta}.  
The parameters resulting from the fit are presented in Table \ref{table:fitParams}. 

\begin{table}[!ht]
\begin{center}
\begin{tabular}{ |p{3cm}||p{3cm}|  }
 \hline
 Fit parameters & Fit results\\
 \hline
 $\omega_\alpha \text{ (meV)}$   & 116.9 $\pm$ 0.1 $\text{meV}$\\
 $\gamma_\alpha^\text{intra}\text{ (meV)}$ & 2.4$\pm$ 0.1 $\text{meV}$ \\
 $\gamma_\beta^\text{intra}\text{ (meV)}$& 9.3 $\pm$ 0.1 $\text{meV}$\\
 $\omega_P\text{ (meV)}$ & 23.4$\pm$ 0.1 $\text{meV}$\\
 \hline  
\end{tabular}
\caption{Parameters returned by the global fit using a quantum master equation model.}%
\label{table:fitParams}
\end{center}
\end{table}

\begin{figure}[ht!]
\ffigbox{%
    \includegraphics[width=\textwidth]{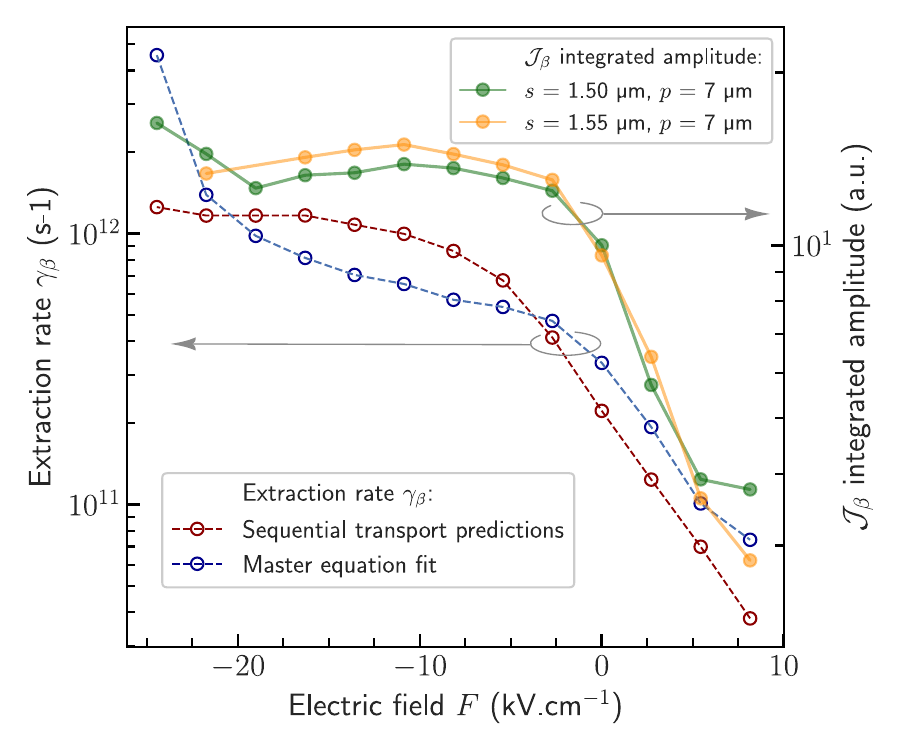}
}{%
  \caption{Left-side scale: extraction rate $\gamma_\beta$ as a function of the applied electric field $F$. Red cross: predicted values computed using a standard sequential transport model. Blue plus sign: values returned by the global fit using a quantum master equation model. Right-side scale: experimental photocurrent integrated amplitude, for two different ($s$, $p$) couples of cavity parameters. }
  \label{fig:gamma_beta}
}
\end{figure}

The returned values are consistent with the previous fits performed with the coupled mode theory in \cite{lagree_direct_2021}. 
In particular, the extraction rate $\gamma_\beta$ as a function of the applied electric field is plotted in Fig.~\ref{fig:gamma_beta} and compared with the values computed through our sequential transport model.  
The right order of magnitude is obtained ($\gamma_\beta < 1$ meV) and the evolution trends are relatively well reproduced ($\gamma_\beta$ decreasing for $F>0$, slope break around $F = -4$ kV.cm$^{-1}$). 
These results on $\gamma_\beta$ are also consistent with the evolution of the integrated amplitude of the spectra (Fig. \ref{fig:gamma_beta}, right-side scale): when the electric field is below $F = -4$ kV.cm$^{-1}$, the electronic cascade is efficiently aligned, and the effective extraction rate $\gamma_\beta$ is high. This leads to a significant photocurrent signal.

The spectrally resolved photocurrent calculated using the parameters returned by the global fit procedure is compared to the experimental data in Fig.~\ref{fig:gamma_beta}, with a quantitative agreement obtained on the set of triplets ($p$, $s$, $F$). 
Two important trends are reproduced as a function of the bias, i.e. as a function of the $\omega_\alpha- \omega_\beta$ alignment: (i) the overall amplitude of the spectra,  and (ii) the relative amplitude inversion between the peaks of the two polaritonic branches.

This study \textit{quantitatively} confirms that the extractor (the electronic cascade of the QCD) and its relative alignment with respect to the ISB transition controls the overall amplitude of the spectra, and also the relative amplitude of the peaks of the polaritonic branches. 
Applying an electric field to the structure enables the selective extraction of excitations from a polaritonic state towards the electronic cascade, while also providing control over the efficiency of this extraction. 
This selective extraction capacity is enabled by the sharp  transfer function and the $2\Omega$ spacing (the Rabi splitting) between the polaritonic peaks: a finer transfer function and a stronger coupling would allow for better selectivity of $\omega_\pm$ polaritons. More details on a QCD transfer function in the strong coupling regime can be found in Appendix \ref{section:transferFunction}.

The good agreement between the experimental data and the theoretical model provides strong evidence that the dark states for both transitions $\alpha$ and $\beta$ do not need to be included in the model to depict an extraction process.
The bright tunnel interaction $\hat{T}_\text{bright}$ and the phenomenological dissipation rate $\gamma_\beta$ from the extractor bright state are sufficient to quantitatively reproduce the experimental measurements.
As previously postulated in \cite{lagree_direct_2021}, this result confirms that the polaritonic nature of the excitation is carried on during the extraction process through the coherent tunnel coupling. 
The extraction is a coherent process, mainly involving the bright states from both $\alpha$ and $\beta$ transitions.  

This model permits however a step forward in the comprehension of the polariton-to-electron process. Chronologically, the early attempts were limited to the observation of a polariton splitting in photo-detection~\cite{dupont_vacuumfield_2003,sapienza_photovoltaic_2007}. A phenomenological transfer function was then introduced in the study of QWIPs operating in strong coupling~\cite{vigneron_quantum_2019}. Recently, the Coupled Mode Theory (CMT) permitted a more rigorous modeling of the transfer function, and an initial indication of direct tunneling into the extractor bright state, with no role for the polaritonic dark states~\cite{lagree_direct_2021}.
The model presented in this paper gets rid of the transfer function - a phenomenological concept - and replaces it with a rigorous tunnel coupling Hamiltonian between the $\alpha$ and $\beta$ transitions, with a complete description of bright and dark states.
The latter do not play a major role for the \textit{polariton extraction} process, but they have a crucial role for \textit{polariton injection}. Our model integrates them, and might constitute a valid vantage point to study electrically injected polariton emitters. More information on the transfer function and the difference between the CMT and the effective density matrix approach can be found in Appendix \ref{section:transferFunction}. 

\section{Implications of the model for electrically pumped polariton emitters: the electron-to-polariton process \label{section:injection}}

%
The validity of the density matrix approach to describe electrical {\it extraction} from optically excited polaritons, motivates to study the implications of these findings on the electrical {\it injection} and subsequent photon emission, represented by the red arrows in Fig.~\ref{fig:system}.
As discussed in Ref.~\cite{de2009quantum}, the main difficulty describing an intersubband emitter operating in the strong light-matter coupling regime lies in the simultaneous description of both optical (bosonic) and electronic (fermionic) excitations.
The injection process fills subband $2$ with fermionic excitations in the form of electrons, while the plasmonic excitations that occupy the $\alpha$ bright state are bosonic.
\rosso{
Working with the full Fermionic Hamiltonian is an arduous task~\cite{de2009quantum}, that could hinder the development of an intuitive understanding of the transport, although very recently a Fermionic approach was successfully used to model QCDs operating in the strong coupling regime~\cite{pisani_electronic_2023}.}

The previous section \ref{sec:extractor} suggests that the bosonization procedure of the extractor, that we employed to describe the extraction process, is a novel and readily interpretable approach for examining the injection process. 
In particular the selection rules for the tunnel Hamiltonian, 
eqs. \eqref{Eq:tunnel_brightbright}-\eqref{Eq:tunnel_darkdark} might prove a powerful tool.
Due to the impossibility of conducting an experimental study resembling the one carried out for QCDs for a detection process, the following discussion will be supported by the quantitative arguments previously presented in section \ref{section:bosonization}. 
Note: the $\beta$ extractor states are now referred to as \textit{injector} states.

An injection process is inherently incoherent because it introduces electrical excitations into an intersubband system through an incoherent external bath of electrons.

The relevant \textit{coherence} here is the one of the 
ISB plasmon\cite{ando_electronic_1982, helm_intersubband_1999,delteil_charge_2012}, that is a collective - and coherent - matter excitation originating from the electronic plasma inside a semiconductor quantum well (QW).
In this respect, an intuitive picture suggests that for an ISB polariton system, the electrical injection process is \textit{not} the reverse of the electrical extraction. In the latter, coherence (induced by light) is destroyed to generate an electrical current, while in the former it appears that coherence must be created.

More formally, in the framework of a bosonized injector, we expect most of the electronic population to be located in the dark states $|B_i^\beta\rangle$ ($i\neq 1$) upon electrical injection. 
Furthermore, to emit light, excitations must be transferred to the plasmonic bright state $|B_1^\alpha\rangle$, which holds the entire oscillator strength of the system.
However, the selection rules \eqref{Eq:tunnel_brightdark} and \eqref{Eq:tunnel_darkbright} are clear: it is impossible for a dark state from the injector to interact with the plasmonic bright state through a tunnel interaction. 
In other words, the primary injection pathway, which would involve direct transfer from the injector states to the bright plasmonic state, can not be taken. 
The bosonized injector formalism confirms that \textit{polaritonic emitters do not operate as reversed polaritonic detectors}. 

In QCDs, the coherence is established through the photonic mode and maintained up to the extractor using both light-matter coupling $\Omega$ and tunnel coupling $\Omega_T$. 
Coherence can also be lost through the irreversible intrasubband scatterings $\gamma_\alpha^\text{intra}$ in the plasmonic mode, although we have demonstrated that it is \textit{not} the main extraction scheme.  However, the extraction process can still take place, since the usual dark-to-dark tunnel interactions are possible (Eq.~\eqref{Eq:tunnel_darkdark}). 
On the contrary, in a LED the injection mechanism is incoherent, and coherence cannot emerge spontaneously during the transport. Additionally, we showed that incoherent (dark) states cannot interact with a coherent (bright) \textit{via} the tunneling Hamiltonian (Eq. \eqref{Eq:tunnel_darkbright}) and (Eq. \eqref{Eq:tunnel_brightdark}). 
As a result, it seems unfeasible to efficiently transfer excitations to the optically active bright state $\alpha$, and thus to the polaritonic states, in the absence of an additional mechanism to generate coherence.
\\
In the case where the electrical injection would be uniform among the $N$ states of $|B_i^\beta\rangle$, light could be emitted since the system would start with excitations in $|B_1^\beta\rangle$, but the expected efficiency would be at most $1/N$, without considering intrasubband decoherence. 

There are however two points that need to be discussed further.
%
%
First, light emission from another kind of polariton states under electrical injection is well documented, namely in exciton-polariton devices \cite{bajoni_polariton_2008,khalifa_electroluminescence_2008,tsintzos_gaas_2008}, with additional reports of polariton lasing under electrical injection \cite{bajoni_polariton_2012,schneider_electrically_2013}.
The key difference is that exciton-polaritons states do not result from a collective matter excitation, but rather from an ensemble of single-particle transitions. 
As a consequence, non-resonant pumping schemes can apply to exciton polaritons, as demonstrated in optical experiments.

Second, several reports of electroluminescence from electrically-injected polariton LEDs exist in the literature.
Some of them clearly determine that thermally assisted emission processes have a major role~\cite{chastanet_surface_2017, askenazi_midinfrared_2017}, but in many other ones simple thermal models cannot explain the data~\cite{sapienza_electrically_2008,delteil_optical_2011,jouy_intersubband_2010,geiser_room_2012}. We can only conjecture possible ways forward to elucidate electrical injection of polaritonic LEDs.
On one hand, one might wonder if the application of the generalized, local Kirchoff~\cite{greffet_light_2018} law to ISB polariton LEDs can shine new light on the electrical injection process, and possibly explain all the existing experimental data in the literature.
%
%
On the other, the problem of electrical excitation of coherent electronic motion - which is essentially the mechanism at play in electrically pumped polariton emitters - is well known from the field of surface plasmon polaritons (SPPs)~\cite{lambe_light_1976,davis_theory_1977,bharadwaj_electrical_2011,parzefall_antenna_2015,kern_electrically_2015,du_highly_2017,qian_efficient_2018}. 
The extremely low efficiency of the electron-to-plasmon and electron-to-photon processes is well known, although recent theoretical works, supported by one experimental finding, have demonstrated that the efficiency could be drastically increased by tailoring the electronic landscape to favor inelastic over elastic tunneling, as long as the electronic coherence is preserved in the process~\cite{uskov_excitation_2016,qian_highly_2021}.

\begin{acknowledgments}
We thank S. De Liberato, J-M Manceau, I. Carusotto, A. Bousseksou for helpful discussions.
We acknowledge financial support from the European Union Future and Emerging Technologies (FET) Grant No. 737017 (MIR-BOSE), and by the French National Research Agency: project SOLID (No. ANR-19-CE24-0003), HISPANID (ANR-17-ASTR-0008-01), and EVEREST (ANR-21-CE24-0021).
\end{acknowledgments}

\clearpage
\appendix

\section{Quantum master equation model for a QCD operating in the strong light-matter coupling regime: parametric study of the impact of the light-matter coupling strength on the transfer function \label{section:transferFunction} }

\begin{figure*}[!ht]
    \centering
    \includegraphics[width=\columnwidth]{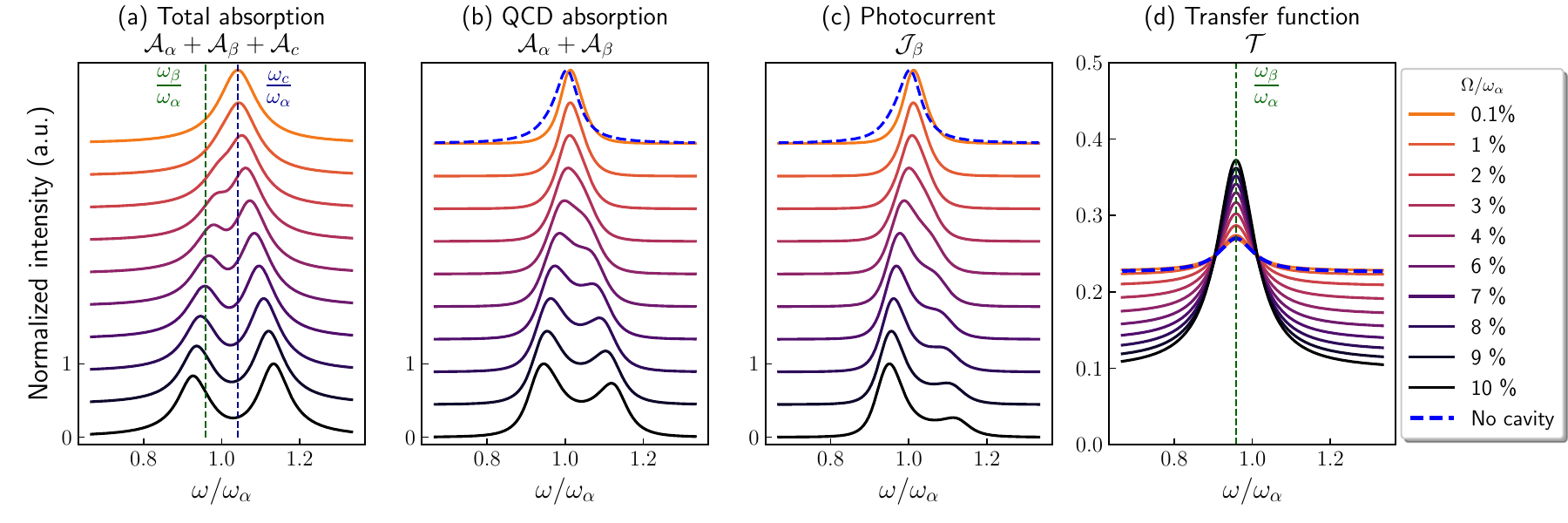}
    \caption{Light-matter coupling $\Omega$ parametric study for the different quantities of interest: [a] Total absorption of the system (ISB absorption, extractor dissipation and cavity absorption) [b] Internal QCD absorption (ISB absorption and extractor dissipation) [c] Photocurrent (extractor dissipation) [d] Transfer function between the extractor dissipation ($\mathcal{A}_\beta := \mathcal{J}_\beta/|s_+|^2$) and the internal QCD absorption ($\mathcal{A}_\alpha + \mathcal{A}_\beta$). The blue dashed line represents an equivalent situation weak coupling situation where the cavity is not included in the model, and the ISB $\alpha$ transition is directly pumped (every other parameters are exactly the same as the strong coupling situations). }
    \label{fig:MasterEq_Parametric}
\end{figure*}

The transfer function between the photocurrent and the total power dissipated inside the QCD ($\mathcal{A}_\text{QCD} = \mathcal{A}_\alpha  + \mathcal{A}_\beta$) is defined as $\mathcal{T}$:
\begin{eqnarray}
\mathcal{T}(\omega) =  \frac{\mathcal{A}_\beta}{\mathcal{A}_\alpha + \mathcal{A}_\beta}
\end{eqnarray}
$\mathcal{T}$ is dependent on the light frequency $\omega$. 

\subsection{Parametric study}

Fig. \ref{fig:MasterEq_Parametric} plots the different quantities $\mathcal{A}_\text{tot}$ ($\mathcal{A}_\text{tot}= \mathcal{A}_\text{QCD} + \mathcal{A}_c$), $\mathcal{A}_\text{QCD}$, $\mathcal{J}_\beta$ and $\mathcal{T}$ computed from the solution of equation \eqref{Eq:QMMEeq}. We impose a realistic situation between the inter and intra-subband dynamics within the QCD such that 90\% of the total broadening is due to the intrasubband scattering:
\begin{eqnarray}
    \gamma_\alpha^\text{intra} + \gamma_\beta^\text{intra} = 0.9 \cdot\gamma_{\alpha \beta} \\
\end{eqnarray}
where $\gamma_{\alpha \beta}= \gamma_\alpha^\text{intra} + \gamma_\beta^\text{intra} + \gamma_\alpha + \gamma_\beta$ represents the total contribution to the broadening from the $\alpha$ and $\beta$ transitions, including intersubband and intrasubband scatterings. 
\rosso{This assumption is equivalent to set $T_{1}\approx 10\cdot T_{2}$, where $T_2$ ($T_1$) are the dephasing (upper state) lifetime, respectively. For a typical mid-IR ISB transition this is verified, as we have $T_1$ of the order of the ps, and $T_2$ of the order of a few hundreds fs.}
The cavity resonance $\omega_c$ and the extractor resonance are also voluntarily mismatched with the ISB transition:
\begin{eqnarray}
    \omega_c = 1.05 \omega_\alpha, ~~~~~~~ \omega_\beta = 0.95 \omega_\alpha
\end{eqnarray}
$\mathcal{A}_\text{tot}$, $\mathcal{A}_\text{QCD}$, $\mathcal{J}_\beta$ and $\mathcal{T}$ are computed for different light-matter coupling amplitudes $\Omega$, up to 10\% of the ISB transition $\omega_\alpha$. 

When the light-matter coupling ratio $\Omega/\omega_\alpha$ increases, the system progressively moves from a weak coupling regime to a strong coupling regime: around the spectral resolution criteria $2\Omega > \gamma_{\alpha\beta}$, we compute the characteristic splitting of the polaritonic peaks, for each spectra $\mathcal{A}_\text{tot}$ (A), $\mathcal{A}_\text{QCD}$ (B) and $\mathcal{J}_T$ (C). The model is able to reproduce the smaller splitting of the QCD absorption (B) compared to the splitting of the total absorption (A) for a same coupling situation $\Omega/\omega_\alpha$, something previously observed in \cite{lagree_direct_2021}. The important novelties that brings the model are found in the transfer function $\mathcal{T}$. In weak coupling (small ratios $\Omega/\omega_\alpha$), the transfer function is almost scalar: it coincides with the transfer function computed in the framework of a QCD that is not inside a cavity. As the ratio $\Omega/\omega_\alpha$ increases, the baseline of the transfer function gradually falls, and the amplitude of its peak increases: increasing $\Omega$ enables the transfer function to reach a Lorentzian shape.

Therefore, in a model where the intra-subband dynamics is explicitly described, the progressive increase of the light-matter coupling allows us to move continuously from a sequential transport in QCDs (flat, quasi-scalar transfer function $\mathcal{T}(\omega)$) to a delocalized description of the transport (sharp, Lorentzian transfer function). Again, when the strong light-matter coupling $\Omega$ is sufficiently intense, the coherent nature of the transport is maintained during the extraction process.

The previous discussion explains the satisfactory description of the photocurrent experimental data produced by the semi-classical CMT obtained in our previous work \cite{lagree_direct_2021}, despite the impossibility in this previous model to describe the intrasubband dynamic. By default, the CMT predicts a sharp Lorentzian transfer function $\mathcal{T}$. While this description is not suited for a weak coupling scenario, where the sequential transport should be described with a scalar transfer function, Fig. \ref{fig:MasterEq_Parametric}-[D] illustrates that it is on the other hand quite adapted to a strong coupling scenario and a delocalized transport scheme. However, being a semi-classical model, the CMT also lacked the ability to distinguish between the inter and intrasubband dynamic which would prevent the distanglement between the spectra broadening and their amplitude. 

\subsection{Tunneling current}

Another quantity of interest is the tunneling current $\mathcal{J}_T$ between the plasmonic mode $\alpha$ and the electronic extraction mode $\beta$. It is defined as:
\begin{eqnarray}
    \mathcal{J}_T = \Omega_T (\braket{ b_\alpha^\dagger b_\beta } - \braket{ b_\alpha b_\beta^\dagger }) 
\end{eqnarray}
Using Eq. \eqref{Eq:QMMEeq} in the low excitation regime, and developing the expressions of the coherences, $\mathcal{J}_T$ can be approximated as:
\begin{eqnarray}
\mathcal{J}_T = & \frac{\displaystyle{2 \Omega_T^2  \gamma_{\alpha \beta}}}{\displaystyle{( \tilde{\omega}_{\alpha} - \omega_\beta )^2 + ( \gamma_{\alpha \beta})^2 }}
                        \left( \braket{b_\alpha^{\dagger}b_\alpha} - \braket{b_\beta^{\dagger}b_\beta} \right)  \nonumber \\
                 &+ \Re\left[ \frac{\displaystyle 2i \Omega_T \Omega}{\displaystyle (\tilde{\omega}_{\alpha} - \omega_\beta  )^2 + ( \gamma_{\alpha \beta})^2 } 
                        \braket{a_c b_\beta^{\dagger}}\right] \label{Eq:J_T}
\end{eqnarray}
where $\gamma_{\alpha \beta}= \gamma_\alpha + \gamma_\alpha^\text{intra} + \gamma_\beta + \gamma_\beta^\text{intra}$ is the sum of the different contributions to the coherences damping inside the QCD. The obtained expression of $\mathcal{J}_T$ in Eq.\eqref{Eq:J_T} is decomposed in two contributions. The fist term is the standard sequential tunnel current \cite{kazarinov1972electric,willenberg2003intersubband} (in its first order expression) which is broadely used for the electronic transport in QCD operating in the weak coupling regime \cite{trinite2011modelling}. It is a semi-classical expression of the current, in the sense that it directly involves the population difference between the modes involved in the tunneling process $  \braket{b_\alpha^{\dagger}b_\alpha} - \braket{b_\beta^{\dagger}b_\beta} $. The second term is a new addition to the tunnel current. It involves the coherences $ \braket{a_c b_\beta^{\dagger}}$ between the cavity and the extractor modes, which could be qualified as long-range coherences (the two modes are only coupled through their mutual coupling to the ISB mode). It thus expresses the system capacity to transport current between modes that are not directly coupled. We will refer to this current as \textit{delocalized current}. 

The amplitude of the delocalized current of Eq.\eqref{Eq:J_T} is controlled by a Lorentzian function and involves the cross product of the couplings $\Omega_T$ (tunnel coupling) and $\Omega$ (light-matter coupling). In the case of a weakly coupled QCD, it is thus expected that the delocalized current is null. Note that it can be numerically checked that the current expression is independent of the considered interface where its computed, thus $\mathcal{J}_T = \mathcal{J}_\beta $.

\subsection{Validity domains of the different models}

\begin{figure}[!ht]
    \centering
    \includegraphics[width=\columnwidth]{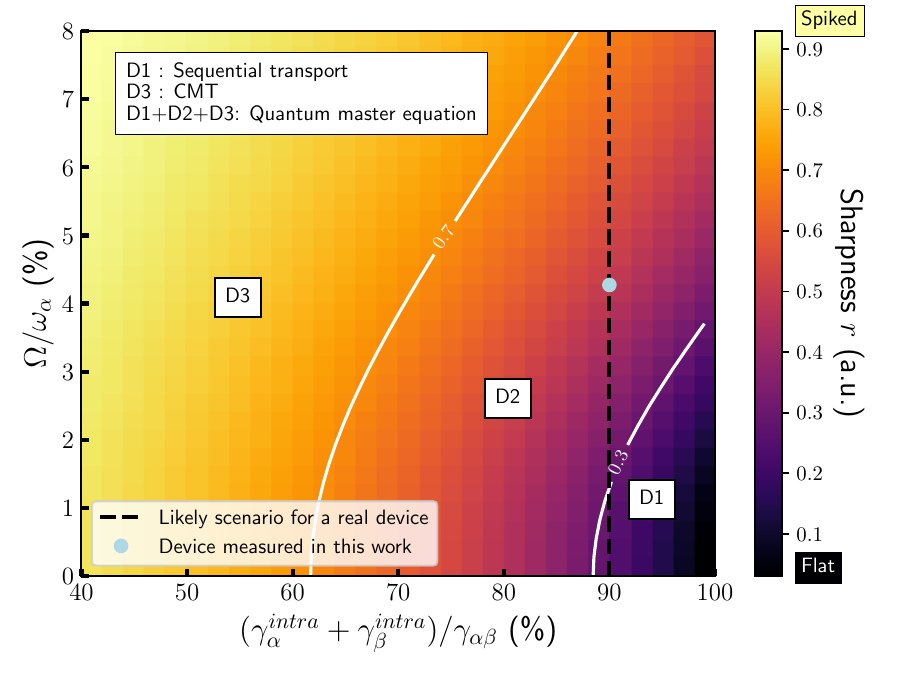}
    \caption{Sharpness of the transfer function $\mathcal{T}$ as a function of the light-matter coupling strength ratio $\Omega/\omega_\alpha$ and the intrasubband scattering predominance $\gamma^\text{intra}$ in both $\alpha$ and $\beta$ transitions with respect to the total scattering $\gamma_{\alpha\beta}$. This map gives intuitive validity domains for all the different physical models considered up to now, but the limits between the domains are arbitrary: a flat transfer function ($r< 0.3$) suggests a sequential transport model, whereas a sharp transfer function ($r> 0.7$) suggests a delocalized transport model, described by the Coupled Mode Theory. The quantum master equation model presented in this article covers the whole domain.}
    \label{fig:MasterEq_Validity}
\end{figure}

To explore the validity domains of the different models introduced here and in our previous work \cite{lagree_direct_2021} (sequential model, CMT model, quantum master equation model), we define a  criterion based on the spectral shape of the transfer function $\mathcal{T}(\omega)$, the sharpness $r$:
\begin{eqnarray}
    r(\Omega, \gamma_\alpha^\text{intra} + \gamma_\beta^\text{intra}) = \frac{\text{Max}\{\mathcal{T}(\omega)\}-\text{Min}\{\mathcal{T}(\omega)\}}{\text{Max}\{\mathcal{T}(\omega)\}}
\end{eqnarray}
$r = 1$ thus indicates that the transfer function $\mathcal{T}(\omega)$ is a sharp Lorentzian function, $r = 0$ indicates that $\mathcal{T}(\omega)$ is a flat scalar function. Fig. \ref{fig:MasterEq_Validity} summarizes the results of the parametric exploration on both the total intrasubband scattering and the light-matter coupling strength. We differentiate three domains D1, D2 and D3:
\begin{enumerate}
    \item Domain D1: sequential transport model, flat scalar transfer function ($\mathcal{T}(\omega)\approx p_E$). This domain is correctly described by the standard thermalized subband model for QCD. \rosso{It corresponds for instance to QCDs operating in the weak-coupling regime}
    \item Domain D3: delocalized transport model, sharp Lorentzian transfer function. This domain is correctly described by the CMT. 
    \item Domain D2: intermediate domain, where transport combines contributions from different sources. D1, D2 and D3 are correctly described by the density matrix formalism of equation \eqref{Eq:QMMEeq} and the capability to distinguish intra and inter-subband dynamic.
\end{enumerate}
\section{Experimental system and QCD bandstructure\label{section:expsystem}}
\begin{center}
\begin{figure}[!ht]
    \includegraphics[width=\columnwidth]{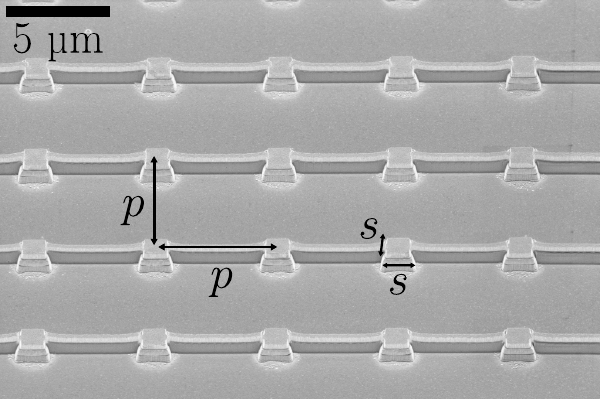}
    \caption{ SEM image of a patch cavity-embedded QCD
detector. $s$ is the patch lateral size and $p$ is the array period.
Patches are electrically connected using gold wires deposited
on a dielectric bridge layer. The active layers, the QCDs, are
embedded between gold layers (Au).}
    \label{fig:patchs}
\end{figure}
\end{center}

\begin{figure}[!ht]
    \centering
    \includegraphics[width=\columnwidth]{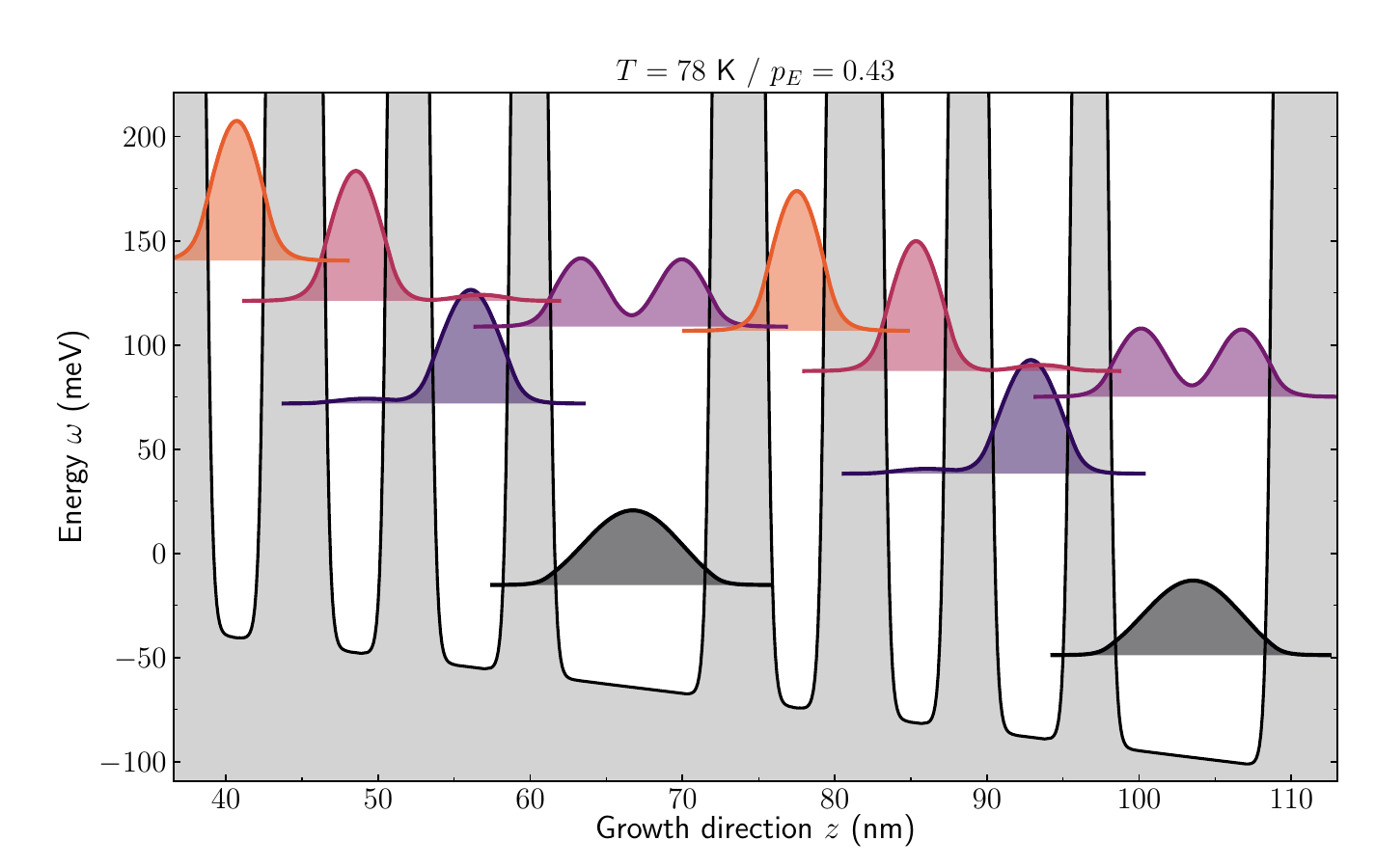}
    \caption{Bandstructure of the QCD used for the photocurrent measurements studied in Fig. \ref{fig:Exp_Photocurrent}. Here the electric field applied on the structure is $F=-10$ kV.cm$^{-1}$. The extraction probability computed for this field and $T=78$ K is $p_E=0.43$ \cite{lagree2022transport}. The potential of the quantum wells is calculated by considering the gradual variation in the composition profile.}
    \label{fig:BandStructure}
\end{figure}

In this section, we present additional information about the samples used in this work. Fig. \ref{fig:patchs} presents a scanning electron microscope (SEM) image of a patch cavity array, and Fig. \ref{fig:BandStructure} presents the bandstructure of the QCD embedded inside the patches. The bandstructure is computed using our sequential transport software \cite{trinite2011modelling}.

\clearpage

\section{Additionnal photocurrent measurements and computational results \label{section:photocurrent_additional}}
In this section, we present additional photocurrent measurements and computational results to supplement the results of Fig. \ref{fig:Exp_Photocurrent}. 
\begin{center}
\begin{figure}[!ht]
    \includegraphics[width=\columnwidth]{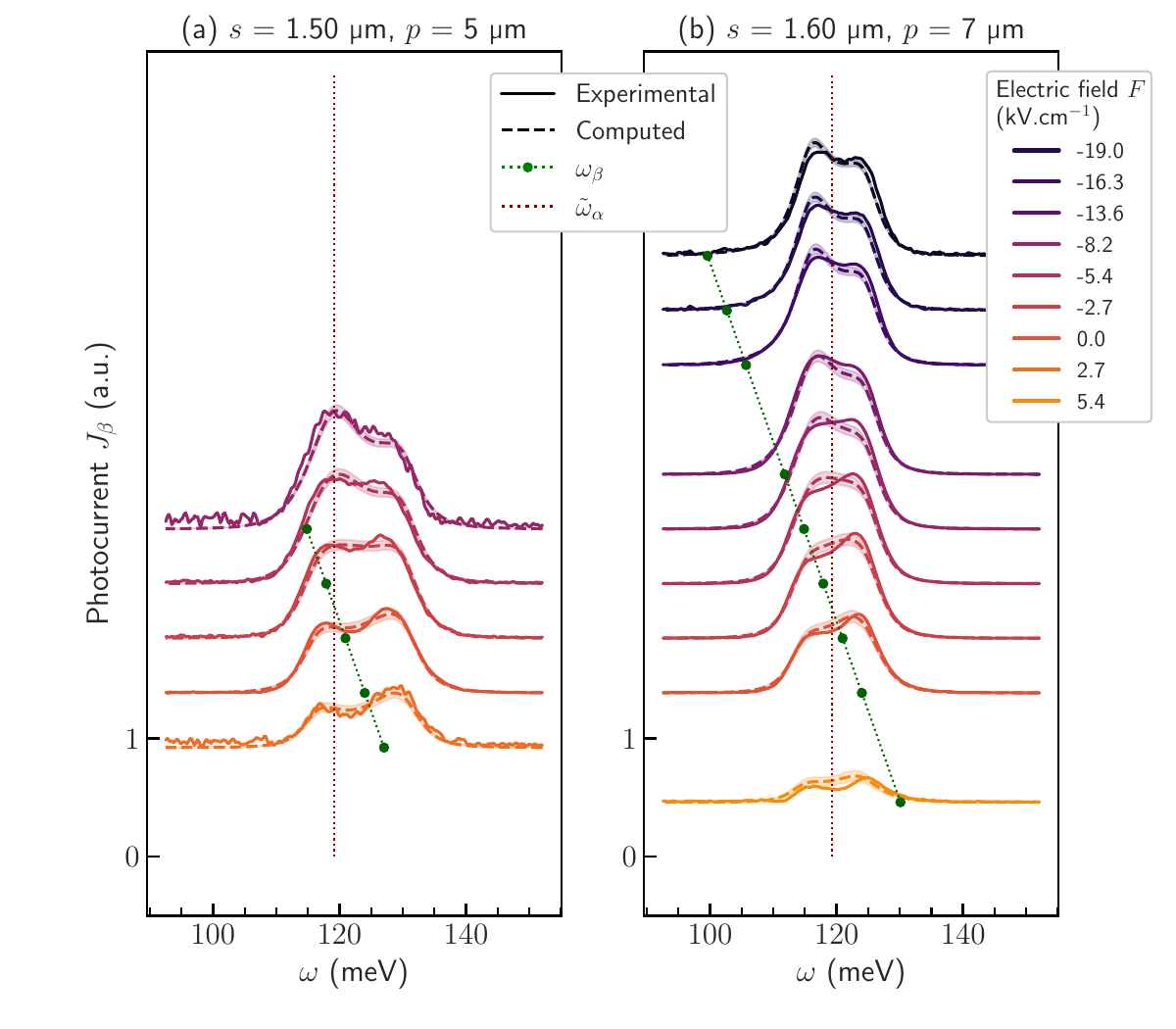}
    \caption{Normalized photocurrent measurements (continuous lines) and quantum master equation global fit (dashed lines), for two cavity geometries [a] $s=1.50$ \textmu m, $p=5$ \textmu m and [b] $s=1.6$ \textmu m, $p=7$ \textmu m. Offsets are added for clarity. Filled areas represent the errors of the fit parameters propagated on the spectra. The extractor frequency $\omega_\beta(F)$, dependent of the electric field $F$, and the plasma-shifted ISB transition $\tilde{\omega}_\alpha$ are both superimposed on the spectra. }
    \label{fig:Exp_Photocurrent_additional}
\end{figure}
\end{center}
%


\end{document}